**Rotational Fluorescence Recovery after Orientational Photobleaching via surface electromagnetic waves on dielectric stacks**

**Francesco Michelotti[1,*], Elisabetta Sepe[1], Agostino Occhicone[1,2], Norbert Danz[3], and Alberto Sinibaldi[1,2]**

[1] *Department of Basic and Applied Science for Engineering, Sapienza University of Rome, Via A. Scarpa 16, 00161 Rome, Italy*

[2] *Italian Institute of Technology, Centre for Life Nano and Neuro Science, Viale Regina Margherita 291, 00161, Roma, Italy*

[3] *Fraunhofer Institute for Applied Optics and Precision Engineering, A.-Einstein-Str. 7, 07745 Jena, Germany*

* *francesco.michelotti@uniroma1.it*

*Abstract* - Protein rotational kinetics are essential for understanding macromolecular behavior in crowded environments, yet measuring these dynamics at solid-liquid interfaces remains a significant challenge due to low signal strengths. Here, we experimentally demonstrate a label-based optical technique for measuring rotational diffusion kinetics using an all-dielectric multilayer stack that sustains both transverse electric and transverse magnetic polarized surface electromagnetic waves. We introduce the concept of Fluorescence Recovery after Orientational Photobleaching, a rotational analogue to the standard translatory fluorescence recovery after photobleaching technique, which utilizes anisotropic photobleaching via resonant transverse electric excitation followed by real-time monitoring of the orientational relaxation towards isotropy. Our ratiometric analysis of the transverse electric and magnetic polarized fluorescence components allows for a distance-independent estimation of the rotational friction coefficient. Applying this method to covalently bound neutravidin, we observe a rotational friction coefficient ($\zeta \approx 5.8 \cdot 10^{-18}\, J \cdot s$) significantly higher than in bulk solutions, highlighting the impact of surface anchoring and molecular crowding. The proposed approach provides a robust, high-sensitivity platform for resolving biomolecular dynamics in complex interfacial environments.

*Introduction* - Protein rotational kinetics can be significantly modulated in crowded environments, such as the intracellular medium, due to steric hindrance and intermolecular interactions. Understanding these dynamics is essential for elucidating protein-protein recognition and behavior under mechanical stress, which may induce denaturation or loss of biological activity. Notably, proteins adsorbed onto surfaces often exhibit long-lived preferred orientations [1–3]. The rotational motion of biomolecules is typically described using the Smoluchowski-Einstein rotational diffusion equation, derived from Brownian drift-diffusion models [4,5].

While nuclear magnetic resonance (NMR) spectroscopy remains the primary tool for investigating internal rotational dynamics and tumbling [6], high-speed atomic force microscopy (AFM) has recently been employed to visualize protein rotation at solid-liquid interfaces, facilitating the design of optimized hybrid biomolecular-inorganic materials [7]. However, detecting molecular orientation within a protein monolayer using conventional transmission Fourier-transform infrared (FTIR) spectroscopy or fluorescence anisotropy (FA) is challenging due to inherently low signal-to-noise ratios. To circumvent these limitations, surface-enhanced (SE) configurations exploiting surface plasmon polariton (SPP) resonances have been implemented for both SE infrared absorption (SEIRA) [8] and SE fluorescence (SEF) [9] spectroscopies. Nevertheless, the use of metal nanostructures and SPP introduces near field effects that near field effects can dominate the results, especially in the SEF case when metal quenching overlays and thus masks orientation dependent effects to be exploited.

In this work, we experimentally demonstrate an optical configuration for measuring the SEF anisotropy of labeled proteins bound at a solid-liquid interface, enabling the estimation of rotational diffusion kinetics. The substrate consists of an inorganic dielectric multilayer stack offering the unique capability to sustain both transverse electric (TE) and transverse magnetic (TM) polarized surface electromagnetic waves (SEW) at the interface [10]. The existence of both TE and TM polarized SEW constitutes an advantage over SPP. Due to the high local density of states (LDOS) associated with SEW, a fluorescent label in proximity to the interface couples to these modes with an efficiency dictated by its transition dipole orientation. In a finite stack, these waves leak into the substrate at discrete angles; their relative intensities can thus be utilized to infer the angular distribution of the labels. The proposed technique relies on the anisotropic photobleaching of initially isotropic labels via resonant TE-polarized SEW excitation, resulting in "angular hole burning." Subsequently, the orientational relaxation towards isotropy is monitored in real-time by detecting both TE- and TM-polarized SEW-coupled fluorescence intensities under attenuated TE illumination. This approach builds upon a theoretical framework for the orientational dynamics of labelled ensembles under angularly selective photobleaching [11] that was also validated experimentally [12].

We herein extend this framework to time-resolved measurements of fluorescence recovery after orientation photobleaching (FROP). We propose FROP as a rotational analogue to fluorescence recovery after photobleaching (FRAP), the latter of which is routinely employed to determine the translational diffusion coefficients of biomolecules in crowded environments. The spectral and polarization-dispersive properties of the SEW-sustaining dielectric stack offer a significantly more compact optical configuration than the standard polarization-resolved FRAP experiments previously reported in the literature [13]. By exploiting the inherent modal selectivity of the multilayer substrate, this approach simplifies the optical path while maintaining high sensitivity to the angular distribution of the transition dipoles.

*Stack design and SEW coupled fluorescence emission* – Herein we study the SEF emission of DyLight 650 organic dye molecules, with absorption and emission bands respectively peaked at $\lambda_{exc}^{max} = 637\,nm$ and $\lambda_{em}^{max} = 650\,nm$, at the surface of purposely designed dielectric stacks described elsewhere [12]. The stacks were deposited by plasma assisted vapor deposition on organic prisms (cyclic olefin copolymer TOPAS 5013 LS, 60 deg corner angle) for operation in water environment and in the visible range under attenuated total internal reflection conditions (Kretschmann–Raether [14]). They were composed by two pairs of bilayers, each constituted by a $Ta_2O_5$ (H) and a $SiO_2$ (L) high and low refractive index layers with the thicknesses $d_H = 120\,nm$ and $d_L = 275\,nm$, respectively. An additional $SiO_2$ layer with thickness $d_L$ was first deposited onto the organic prism to improve the reliability of the stack. The stack was topped with a $SiO_2/TiO_2$ bilayer, with thickness of both layers $d_T = 20\,nm$. Cross sections of test stacks deposited on reference cover slip substrates were inspected by field-emission scanning electron microscopy and the thicknesses of the layers matched within the uncertainty of the design values. The complex refractive indices at λ = 670 nm of $Ta_2O_5$, $SiO_2$, and $TiO_2$were $\tilde{n}_H = 2.160 + i5 \times 10^{-5}$, $\tilde{n}_L = 1.474 + i5 \times 10^{-6}$, and $\tilde{n}_{TiO_2} = 2.28 + i1.8 \times 10^{-3}$, respectively. The prism and water/solution refractive indices at λ were $n_{pr} = 1.526$ and $n_{sol} = 1.328$, respectively. A two-component flow cell, consisting of a hard polymer cover and an elastomer film that defined microchannel, was used to inject aqueous fluids in a channel at the surface of the stack.

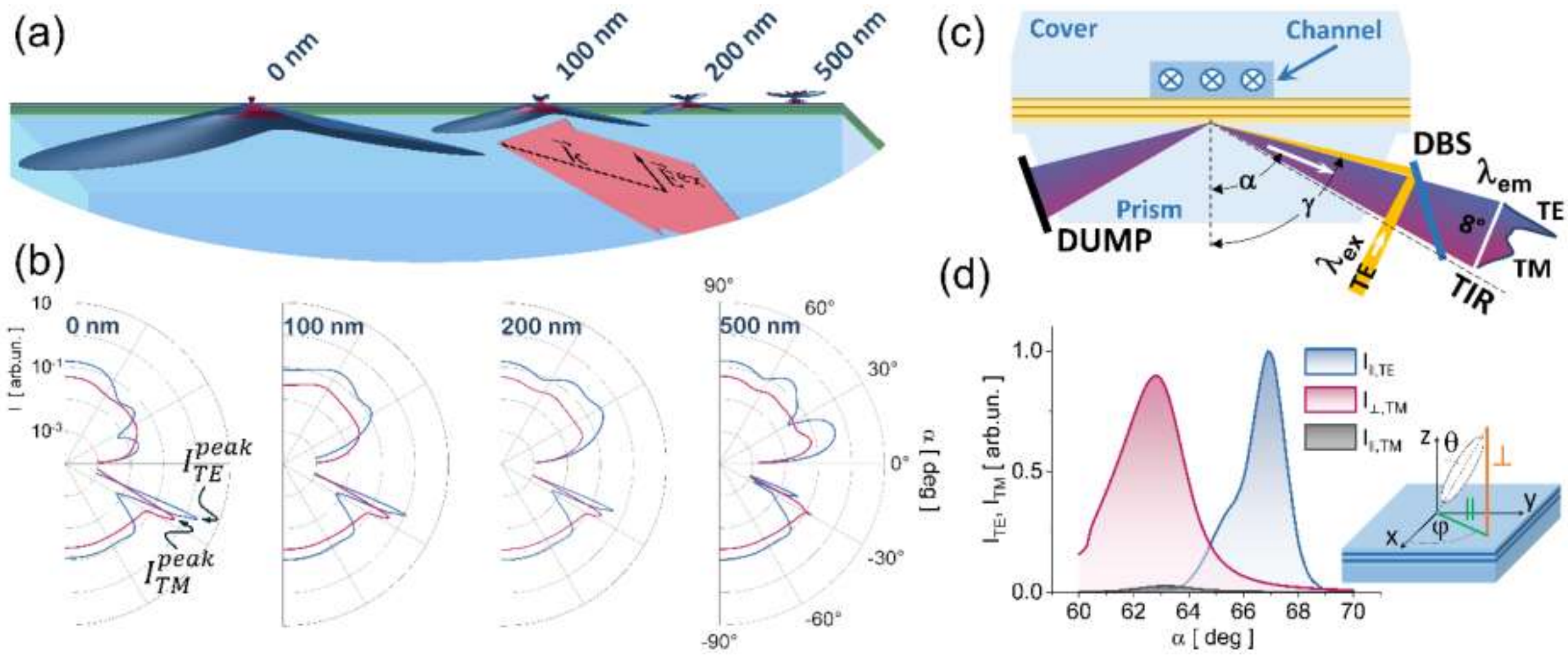

***Figure 1***. (a) TE (blue) and TM (red) polarized angular fluorescence emission patterns of an isotropically oriented DyLight 650 ensemble at several distances from the surface of the specific dielectric stack described in the text. The TE emission cones were purposely cut to show the internal TM emission cones. (b) Cross-sections of (a) along the incidence plane (same color coding, log scale). (c) Sketch of the experimental configuration used to resonantly excite at $\lambda_{exc} = 637\ nm$ the chromophores at the surface of the stack and to collect their angularly dispersed fluorescence in the $\lambda_{em} \in [640{,}730]\ nm$ range. (d) Calculated TE and TM polarized contributions to the fluorescence emission of DyLight 650 dipoles laying either parallel ($I_{\parallel,TE}$ blue, $I_{\parallel,TM}$ red) or perpendicular ($I_{\perp,TM}$ black) to the stack surface.

In Figure 1a we show the stationary fluorescence emission of an ensemble of isotropically oriented identical DyLight 650 emitters, with orientational distribution function $f_0(\theta,\varphi) = 1/4\pi$, placed in proximity of the surface of the stack, for several different distances $d$. The 3D patterns were calculated by means of a rigorous Greens' functions approach [11,15], when neglecting depolarization during the excitation lifetime [16] and photobleaching (PB) [17]. The dye was modeled as a rod-like emitter with aligned excitation ($\mu_{exc}$) and emission ($\mu_{em}$) dipole moments and the patterns were integrated over the emission spectrum peaked around $\lambda_{em}$. Excitation under resonant SEW coupled TE illumination at $\lambda_{exc}$, in the configuration sketched in Figure 1c, sets an anisotropic population of excited emitters that then fluoresce coupling to the SEW leaky modes of the stack, which finally radiate into the prism. Dipoles with $\mu_{em}$ aligned along the direction either perpendicular ($\perp$) or parallel ($\parallel$) to the stack surface contribute differently to the TE and TM polarized SEF emission, as shown in Figure 1d in which we plot their normalized emission. The 3D patterns were calculated considering the TE excitation, decomposing $\mu_{em}$ along the $\perp$ and $\parallel$ directions and averaging the respective contributions to the radiation patterns over the $f_0(\theta,\varphi)$ distribution [11]. The distance $d$ governs coupling to the SEW, which shows a different strength for the TE (blue) and TM (red) polarization, owing to their different LDOS transverse distribution. Only the emitters that are closer than the penetration depth of the SEW into the liquid medium, which is about 110 nm for the TE polarization [11], efficiently couple to such modes and radiate into the prism.

The 1D angular emission curves shown in Figure 1b are the sections of the 3D patterns within the incidence plane, with the TE and TM SEW coupled emission laying beyond the TIR edge. Their peak intensities $I_{TE}^{peak}$ and $I_{TM}^{peak}$ increase for smaller $d$, with a ratio $R = I_{TE}^{peak}/I_{TM}^{peak}$ that also depends on $d$ [11].

*Fluorescence recovery after orientational photobleaching (FROP)* – PB can significantly drive the orientational distribution $f(\theta,\varphi,t)$ of the ensemble of active emitters off the isotropic $f_0(\theta,\varphi)$. In a previous article [11], we modelled the effect of PB by means of the modified Smoluchowski-Einstein roto-diffusional equation for rod-like molecules [18]:

$$\frac{\partial f(\theta,\varphi,t)}{\partial t} = \frac{k_B T}{\zeta}\frac{1}{sin\theta}\frac{\partial}{\partial\theta}\left(sin\theta\frac{\partial f(\theta,\varphi,t)}{\partial\theta}\right) + \frac{k_B T}{\zeta}\frac{1}{sin^2\theta}\frac{\partial^2 f(\theta,\varphi,t)}{\partial\varphi^2} - \sigma_{PB} I_{exc} f(\theta,\varphi,t) sin^2\theta sin^2\varphi \tag{1}$$

where $k_B T$ is the thermal energy, $\zeta$ is a friction coefficient between the molecule and the host medium, $I_{exc}$ is the TE polarized illumination intensity and $\sigma_{PB}$ is the PB cross-section. We consider irreversible PB and $Q(t) = \int f(\theta,\varphi,t) d\Omega < 1$ is the percentage of active emitters at the time *t*. Here we also assume that $\zeta$ is sufficiently large to avoid depolarization during the emitter's lifetime in the excited state (2-10 ns range) and small enough to observe depolarization on the timescale of cw fluorescence experiments (hundreds of seconds). Eq. (1) can be solved by expanding $f(\theta,\varphi,t)$ over the Laplace spherical harmonics $Y_l^m(\theta,\varphi)$, which are solutions of the pure rotational diffusion equation [18], with time dependent coefficients $A_l^m(t)$:

$$f(\theta,\varphi,t) = \sum_{l=0}^{\infty}\sum_{m=-l}^{m=l} A_l^m(t) Y_l^m(\theta,\varphi) \tag{2}$$

where:

$$Y_l^m(\theta,\varphi) = a_{lm}\, e^{im\varphi}\, P_l^m(cos\theta) \tag{3}$$

$$a_{lm} = \frac{1}{\sqrt{2\pi}}(-1)^m\sqrt{\frac{(2l+1)(l-m)!}{2(l+m)!}} \tag{4}$$

and $P_l^m(cos\theta)$ are the associated Legendre polynomials [19].

Under this approach, Eq. (1) reduces to a set of coupled differential equations for the $A_l^m(t)$:

$$\frac{d\boldsymbol{A}(t)}{dt} = \boldsymbol{H}\cdot\boldsymbol{A}(t) = \left(\boldsymbol{H}^{(\boldsymbol{PB})} + \boldsymbol{H}^{(\boldsymbol{DEP})}\right)\cdot\boldsymbol{A}(t) \tag{5}$$

where:

$$\boldsymbol{A}(t) = \left(A_0^0(t), A_1^{-1}(t), A_1^0(t), A_1^1(t), \dots, A_N^N(t)\right) \tag{6}$$

N is the maximum order of the expansion and the matrix operators $\boldsymbol{H}^{(\boldsymbol{PB})}$ and $\boldsymbol{H}^{(\boldsymbol{DEP})}$ [11] govern PB and depolarization, respectively.

The solution of the Eq. (5) is:

$$\boldsymbol{A}(t) = C_1 e^{\lambda_1 t}\boldsymbol{u_1} + C_2 e^{\lambda_2 t}\boldsymbol{u_2} \ldots + C_M e^{\lambda_M t}\boldsymbol{u_M} \quad (7)$$

where $\lambda_i$ and $\boldsymbol{u_i}$ are the eigenvalues and eigenvectors of $\boldsymbol{H}$ and the constants $C_i$ depend on the initial conditions $A_l^m(0)$. Once the $A_l^m(t)$ are found, one can calculate $f(\theta, \varphi, t)$ and determine analytically the time dependent emission patterns and the temporal evolution of the $R(t)$ parameter [11]. If the fluorescence detection is carried out in the incidence plane, we showed that only few $A_l^m(t)$ are non-zero, namely those with even $l$ up to 4 and with even $m$.

Such a theoretical model was used to quantitatively fit the results of experimental assays [12] in which the $R(t)$ parameter was measured in real-time under cw SEW coupled TE illumination, under the conditions in which the PB process was prevailing over depolarization, therefore leading to an anisotropic emitter orientation and finally to complete PB. We point out that the $R(t)$ analysis is ratiometric [20] and therefore independent on the number of active emitters and on their distance $d$ from the stack surface. In [12], to fit the data, it was necessary to tune both the $\frac{k_B T}{\zeta}$ and $\sigma_{PB} I_{exc}$ free parameters. The latter is experimentally very uncertain, since the linear model we assumed for PB with a constant cross-section could be oversimplified [13] and the SEW intensity at the stack surface depends on a number of parameters, such as the coupling efficiency and the resonance width, which are not easily controllable. This makes that the experimental value found $\frac{k_B T}{\zeta}$ could be affected by a large error.

The latter observation is the reason why it would be desirable to operate with just one free parameter and led us to theoretically propose a measurement procedure in which the orientational distribution of an emitter ensemble is first driven off-isotropy by a strong TE polarized illumination and angularly selective PB until the time $t_{PB}$, and then the relaxation of the distribution towards isotropy is measured by a weak TE polarized illumination and fluorescence detection [11]. During relaxation, $f(\theta, \varphi, t)$ follows Eq. (1) with $\boldsymbol{H}^{(\boldsymbol{PB})} = 0$:

$$\frac{dA(t)}{dt} = \boldsymbol{H}^{(\boldsymbol{DEP})} \cdot \boldsymbol{A}(t) \quad (8)$$

Since $\boldsymbol{H}^{(\boldsymbol{DEP})}$ is diagonal [11] with:

$$H_{l(l+1)+m,l(l+1)+m}^{(DEP)} = -\, k_B T/\varsigma\, l(l+1) = -D_l \quad (9)$$

we get:

$$A_l^m(t) = A_l^m(t_{pb}) e^{-D_l(t-t_{pb})} \tag{10}$$

Since only $A_0^0$, $A_2^m(t)$ and $A_4^m(t)$ are non-zero, the recovery of $R(t)$ is governed just by $D_2 = 6k_BT/\zeta$ and $D_4 = 20k_BT/\zeta$. Such last result suggests that by performing FROP experiments and monitoring the recovery of the ratiometric quantity $R(t)$ [11], one can directly access to the rotational diffusion coefficient of any fluorescent molecule bound in proximity of the 1DPC surface with the spatial resolution provided by the SEW evanescent tail.

In Figure 2 we show the temporal evolution of the normalized $R(t)/R(0)$ parameter during a simulated FROP experiment. The curves were obtained by analytically solving the Eq. (1) for a population of DyLight 650 emitters at the surface ($d = 0$) of the 1DPC. During the PB phase, under strong excitation conditions until $t_{PB} = 60\ s$ with $\sigma_{PB} I_{exc}$ values between $5\ ks^{-1}$ and $100\ ks^{-1}$, $R(t)$ decreases as a result of the angular hole burning of the distribution of the active emitters $f(\theta, \varphi, t)$ along the TE illumination direction. The first four curves were obtained for $k_BT/\zeta = 1\ ks^{-1}$, the two last curves for $k_BT/\zeta = 0.5\ ks^{-1}$ and $k_BT/\zeta = 0.1\ ks^{-1}$, respectively. The more PB prevails on rotational diffusion, the more $R(t)$ is driven out of the isotropy value and the more $f(\theta, \varphi, t)$ is distorted, as also shown by the $f(\theta, \varphi, t)$ surface plots insets for the first four curves at $t_{PB}$. During the orientational diffusion phase ($t > t_{PB}$), the $R(t)/R(0)$ recovers under low intensity illumination ($\sigma_{PB} I_{exc} = 0.05\ ks^{-1}$) towards unity with a different time constant depending on the $k_BT/\zeta$ value. For the first four curves complete relaxation is achieved during the simulation time and $f(\theta, \varphi, t)$ recovers to the isotropic condition, with different residual values of $Q(t) < 1$, due to irreversible PB. For the two last curves the relaxation is not complete during the simulation time and the residual $f(\theta, \varphi, t)$ is still distorted. Simulations carried out under a wide range of values of $k_BT/\zeta$ show that the relaxation is well described by a single exponential with $D_2 = 6k_BT/\zeta$ and that the $D_4 = 20k_BT/\zeta$ term can be neglected.

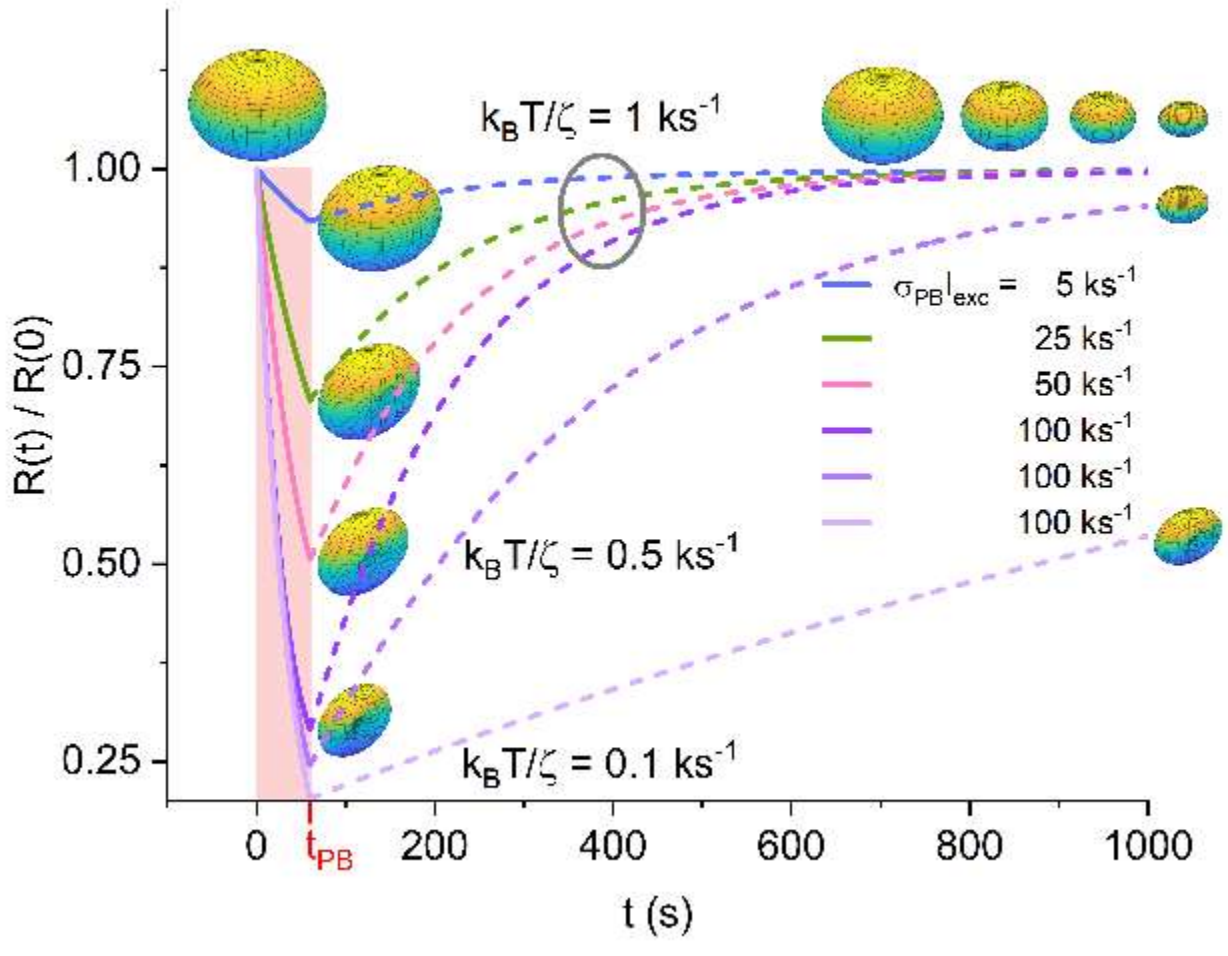

***Figure 2***. Analytical calculation of the temporal evolution of $R(t) = I_{TE}^{peak}/I_{TM}^{peak}$ normalized to $R(0)$. During the PB phase (red region) for $t \in [0, t_{PB} = 60]\ s$ we set $\sigma_{PB} I_{exc} \in [5, 100]\ ks^{-1}$ and during the recovery phase $\sigma_{PB} I_{exc} = 0.05\ ks^{-1}$. The curves were calculated for several different values of $k_B T/\zeta \in [0.1, 1]\ ks^{-1}$. The inset figures are the $f(\theta, \varphi, t)$ distributions calculated at $t = 0$, $t = t_{PB}$, and $t = 1000\ s$, for the different conditions.

*Experimental procedure and results* - The surface of pristine biochips constituted by dielectric stacks deposited onto organic chips were chemically and biochemically modified, according to the procedures described below. Sulfuric acid (95–98%), hydrogen peroxide (30% in H2O), (3-aminopropyl)-triethoxysilane (APTES, 99%), ethanol (99.8%), glutaraldehyde solution (GAH, grade I, 50% in H2O), sodium bicarbonate (99.7%), sodium cyanoborohydride ($NaCNBH_3$, 95%), NeutrAvidin protein, DyLight 650 (NAv650), and Dulbecco's phosphate buffered saline 10×, (D-PBS 10×, 100 mM) were purchased from Sigma-Aldrich and used as received.

Neutravidin, a deglycosylated form of avidin, retains high biotin-binding affinity while exhibiting reduced nonspecific interactions. It has a molecular weight of ~60 kDa and a near-neutral isoelectric point (pI ≈ 6.3), features that enhance its performance in biochemical assays relative to native avidin. Conjugation with DyLight 650 is typically controlled to preserve binding functionality, yielding a degree of labeling of 1–3 fluorophores per protein. This range ensures efficient fluorescence while minimizing steric and structural perturbations that could impair biotin binding [21].

The first step of the chemical surface functionalization was a piranha treatment to increase the hydroxyl groups density and remove contaminants. Then the stacks were immersed in 2% APTES in a mixture of ethanol/water (95/5 v/v) for 1 h at ambient temperature, followed by sonication, washing in ethanol, and soft-baking on a hot plate at 60 °C for 1 h. Finally, they were immersed in 2% GAH in bicarbonate buffer for 1 h at ambient temperature adding $NaCNBH_3$, followed by sonication and washing in deionized water.

In the experiments, a biochip was topped with microfluidics and mounted on the experimental setup. The fluidic channel was primed with D-PBS 1× and the fluorescence background was acquired. Then, 160 µL of NAv650 at a concentration of 0.5 mg/mL at a flow rate of 1.35 µL/s followed by a washing step with 1 mL in D-PBS 1× at a flow rate of 3.59 µL/s. Washing was repeated twice with a total volume of 2 mL. The biochip was kept in dark after protein binding.

Before starting a measurement, the fluorescence excitation laser incidence angle $\gamma$ was tuned to resonantly excite a SEW at $\lambda_{exc}$ and the average intensity $I_{exc}^{PB} = I_0$ was let to stabilize, with a closed shutter along its path. During the experiments, the biochip temperature was kept constant at 30 °C. At $t = 0$, the shutter was opened and the angular fluorescence spectrum was continuously collected (with an integration time of $5 \times 10^{-4}$ s). During such a PB phase, an exponential decay of the TE and TM peak intensities , $I_{TE}^{peak}$ and $I_{TM}^{peak}$ , is observed until $t_{PB} = 60\ s$, when the shutter was closed again, as shown in Figure 3. After PB, we inserted a neutral density filter to decrease $I_{exc}$ by a factor 70 and changed the integration time to $5{\times}10^{-2}$ s. The shutter was closed and opened periodically for about 2000 s, keeping alternately the biochip under dark ($I_{exc} = 0$) and attenuated illumination ($I_{exc}^{REC} = I_{exc}^{PB}/70 = I_0/70$) conditions for 15 s and 3 s, respectively, as shown in the inset of Figure 3. Under such conditions, the PB process was slowed down by a factor 420 compared to the PB phase and rotational diffusion prevailed, giving rise to fluorescence recovery apparent as slow increase of $I_{TE}^{peak}$ and $I_{TM}^{peak}$ over observation time, as shown in Figure 3.

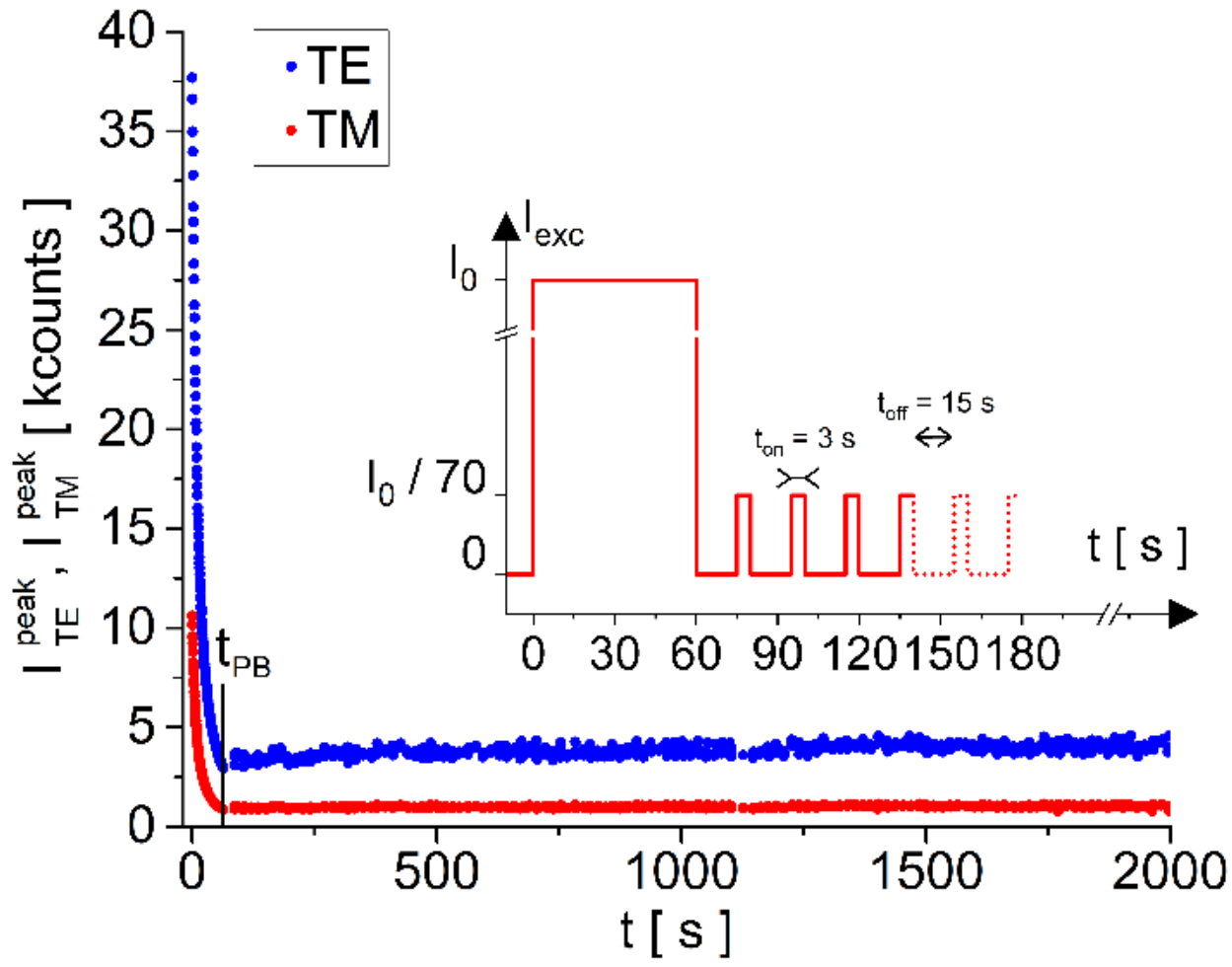

Figure 3. Measured temporal dependencies of both the $I_{TE}^{peak}$ and $I_{TM}^{peak}$ peak intensities during a PB for $t \in [0, t_{PB} = 50]\ s$ and subsequent fluorescence recovery experiment. The inset illustrates the temporal illumination conditions.

In the inset of Figure 4, we show the magnified time dependencies of $I_{TE}^{peak}$ and $I_{TM}^{peak}$ during the recovery phase after $t_{PB}$. As expected, the intensity of the TE polarized component recovers more than the TM, since it was driven more out of equilibrium. The normalized ratio $R(t)/R(0)$ is plotted all over the experiment. When opening the shutter, the ratio starts decreasing due to PB prevailing on rotational diffusion, as predicted by the simulations reported in Figure 2. After $t_{PB}$, a single exponential recovery is observed towards isotropy. However, we did not observe a complete recovery, that we attribute to a population of proteins that were bound to the surface of the biochip in multiple points and that suffered a strong limitation to their orientational mobility. Apart the first initial growth, the measurement could be fitted by the predictions of the Eq. (1), when assuming $\sigma_{PB} I_{exc}^{PB} = 0.0124\ ks^{-1}$, $k_B T/\zeta = 0.72\ ks^{-1}$, $\sigma_{PB} I_{exc}^{REC} = 0.05\ ks^{-1}$ and assuming that 17% of the population of proteins at the surface completely lacked orientational mobility.

From the fit of $R(t)/R(0)$ in Figure 3, one can retrieve the value of the rotational friction coefficient of the mobile fraction of proteins at 30 °C, which is about $\zeta = 5.8 \cdot 10^{-18}\, J \cdot s$. Such a value can be compared to those reported in the literature for several different conditions. For a free neutravidin protein with a molecular weight of 60 kDa, when approximating a sphere with hydrodynamic radius $r \approx 3.7\, nm$ dissolved in water with viscosity $\eta \approx 0.8 \cdot 10^{-3} Pa \cdot s$ at 30 °C, the predicted value for the rotational friction coefficient according to the Debye-Stokes-Einstein relation with a general slip boundary condition [22] is $\zeta = 8\pi\eta r^3 \approx 10^{-27} J \cdot s$. This latter value is notably low, consistent with the unrestricted rotational freedom of the system. Recently, Zhang et al. [7] found that recombinant human Mica 6 proteins adsorbed at a water/solid interface show a rotational friction coefficient for in plane rotations in the range of $\zeta \approx 10^{-19} J \cdot s$. It is therefore likely that covalently bound proteins, such as neutravidin tethered via GAH to a surface in the present work, show an even larger friction coefficient as found in our experiments.

It is worth pointing out that, if at the end of the fluorescence recovery phase the excitation intensity is increased again to $I_{exc}^{PB} = I_0$, the system can still be driven out of equilibrium, as shown in Figure 4. Even though in Figure 3 a strong PB is observed, there is still a residual population of active emitters that can be used for the FROP analysis with a large dynamic range.

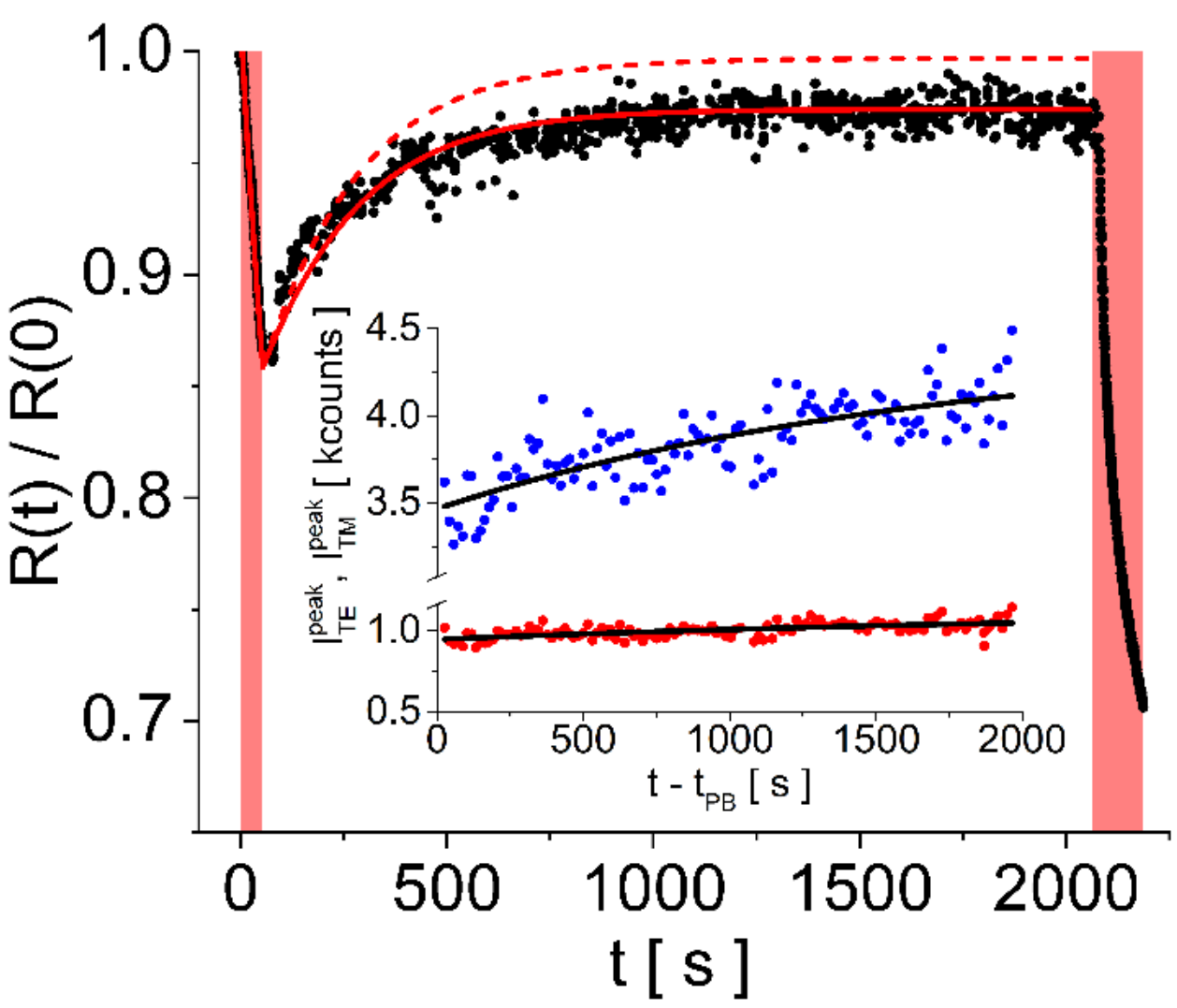

***Figure 4***. Temporal dependency of the normalized $R(t)/R(0)$ extracted from the experimental curves shown in Figure 3. The fits were obtained by the Eq. (1), when assuming that all residual active dipoles are either free to rotate (dashed) or partially immobilized (solid). The inset shows a magnified plot of the $I_{TE}^{peak}$ and $I_{TM}^{peak}$ signals shown in figure 3, limited to the fluorescence recovery phase. The last part of the measurement (second red region) was acquired under PB conditions.

*Conclusions* - In summary, we have demonstrated a novel optical configuration that exploits the unique properties of dielectric multilayer stacks to investigate the rotational diffusion kinetics of proteins at solid-liquid interfaces. By harnessing the simultaneous excitation of TE and TM polarized surface electromagnetic waves (SEW), we introduced and validated the Fluorescence Recovery after Orientational Photobleaching (FROP) technique. This approach serves as a rotational analogue to the well-established FRAP, providing high surface sensitivity due to the large local density of states and the evanescent nature of the SEW. Our experimental results on covalently bound neutravidin molecules reveal a significant increase in the rotational friction coefficient compared to bulk solution values, highlighting the restrictive role of surface anchoring and local crowding on molecular mobility. Notably, the ratiometric nature of the $R(t)$ analysis ensures that the measurements are intrinsically independent of the fluorophore concentration and the precise emitter distance from the surface, offering a robust tool for quantitative biophysical studies. Beyond the specific case of protein adsorption, this SEW-based FROP technique opens new avenues for the real-time monitoring of biomolecular dynamics in complex environments, such as lipid membranes or artificial cellular interfaces. The ability to resolve orientational relaxation with high spatial resolution and sensitivity suggests that this method could become a benchmark for designing optimized hybrid biomolecular–inorganic materials and for fundamental studies of protein behavior under mechanical or steric stress.

*Acknowledgements* - The present work was funded by the Italian Ministry of Research, under the complementary actions to the NRRP "D34health - Digital Driven Diagnostics, prognostics and therapeutics for sustainable Health care" Grant (# PNC0000001, project code B53C22006120001). The authors wish to thank P.Munzert for the multilayers deposition.

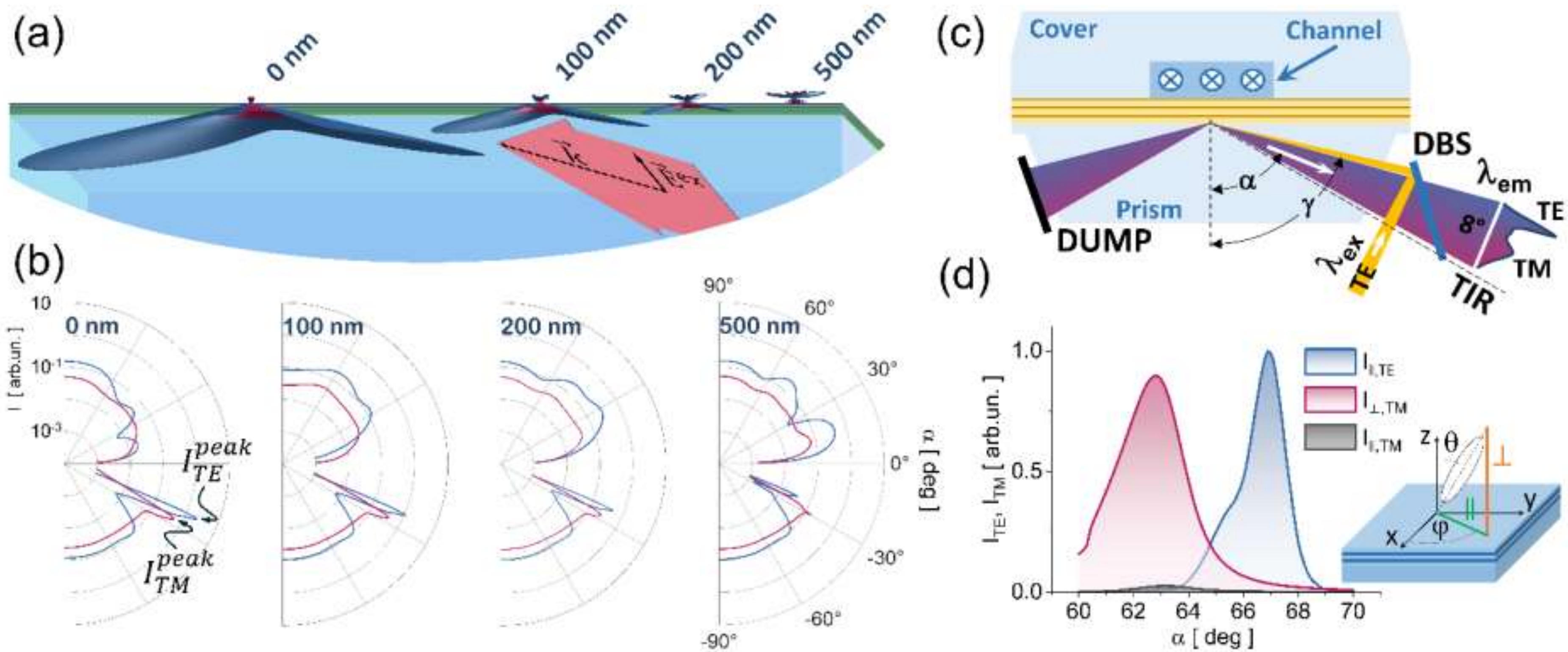

***Figure 1***. (a) TE (blue) and TM (red) polarized angular fluorescence emission patterns of an isotropically oriented DyLight 650 ensemble at several distances from the surface of the specific dielectric stack described in the text. The TE emission cones were purposely cut to show the internal TM emission cones. (b) Cross-sections of (a) along the incidence plane (same color coding, log scale). (c) Sketch of the experimental configuration used to resonantly excite at $\lambda_{exc} = 637\ nm$ the chromophores at the surface of the stack and to collect their angularly dispersed fluorescence in the $\lambda_{em} \in [640{,}730]\ nm$ range. (d) Calculated TE and TM polarized contributions to the fluorescence emission of DyLight 650 dipoles laying either parallel ($I_{\parallel,TE}$ blue, $I_{\parallel,TM}$ red) or perpendicular ($I_{\perp,TM}$ black) to the stack surface.

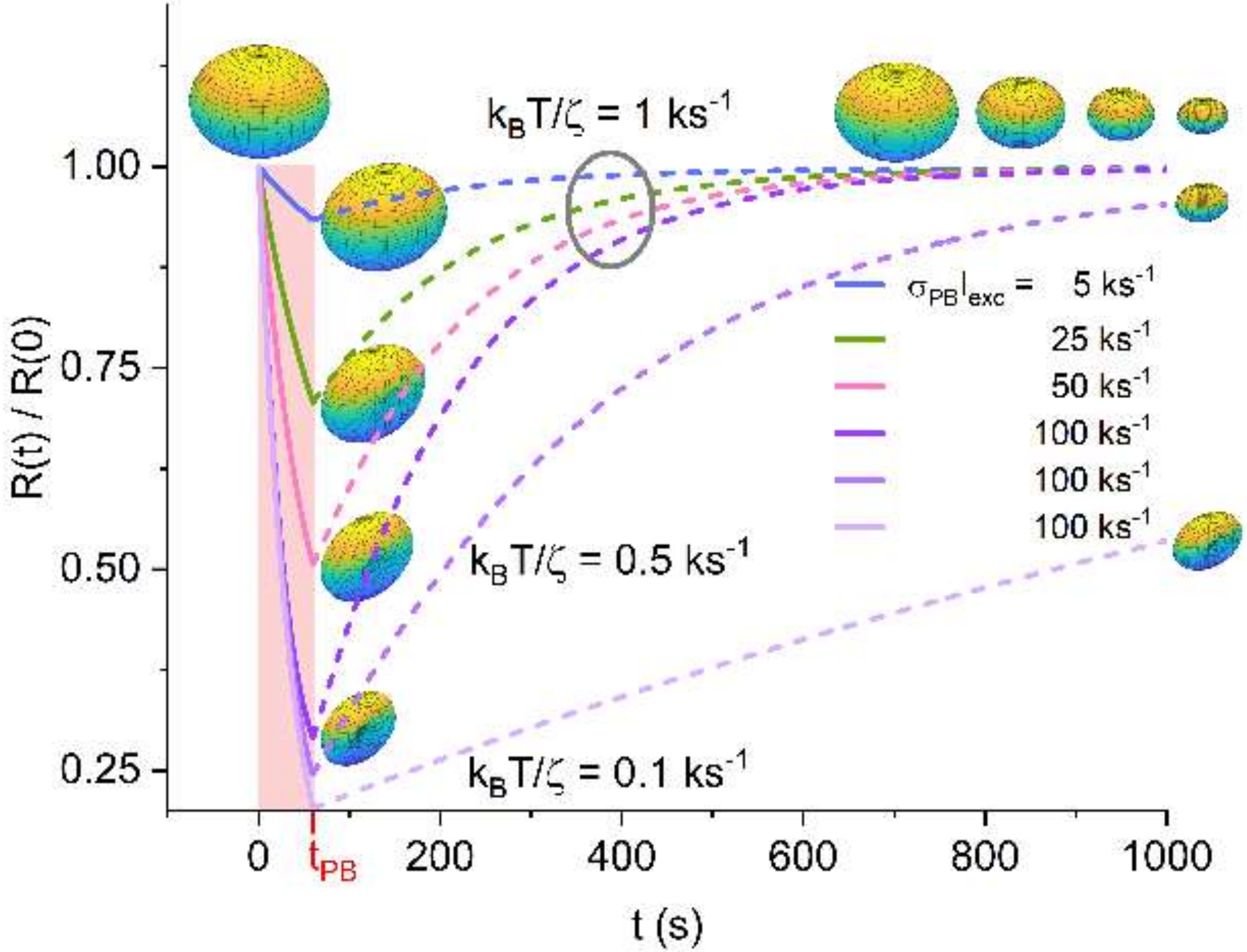

***Figure 2***. Analytical calculation of the temporal evolution of $R(t) = I_{TE}^{peak}/I_{TM}^{peak}$ normalized to $R(0)$. During the PB phase (red region) for $t \in [0, t_{PB} = 60]\ s$ we set $\sigma_{PB} I_{exc} \in [5, 100]\ ks^{-1}$ and during the recovery phase $\sigma_{PB} I_{exc} = 0.05\ ks^{-1}$. The curves were calculated for several different values of $k_B T/\zeta \in [0.1, 1]\ ks^{-1}$. The inset figures are the $f(\theta, \varphi, t)$ distributions calculated at $t = 0$, $t = t_{PB}$, and $t = 1000\ s$, for the different conditions.

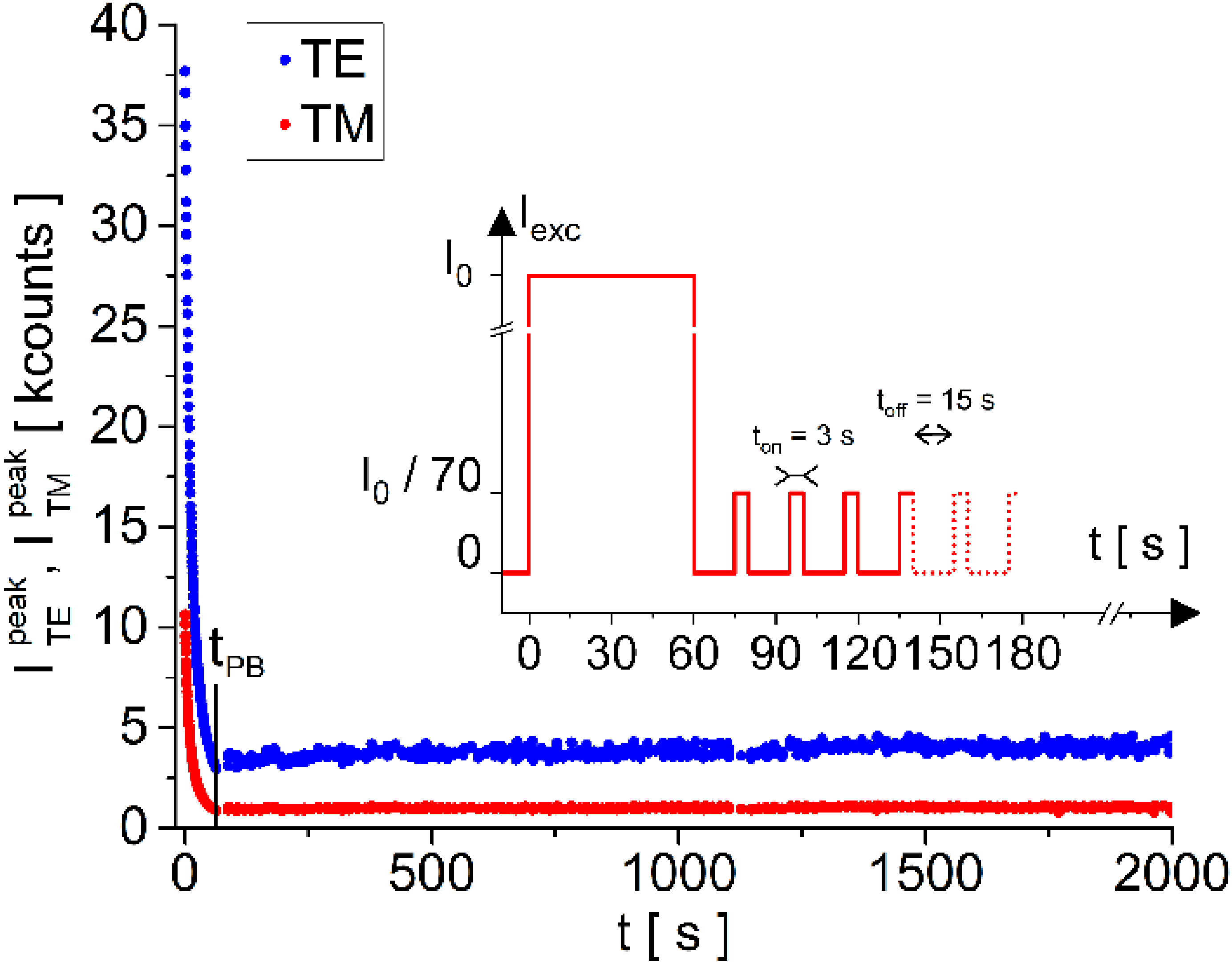

Figure 3. Measured temporal dependencies of both the $I_{TE}^{peak}$ and $I_{TM}^{peak}$ peak intensities during a PB for $t \in [0, t_{PB} = 50]\ s$ and subsequent fluorescence recovery experiment. The inset illustrates the temporal illumination conditions.

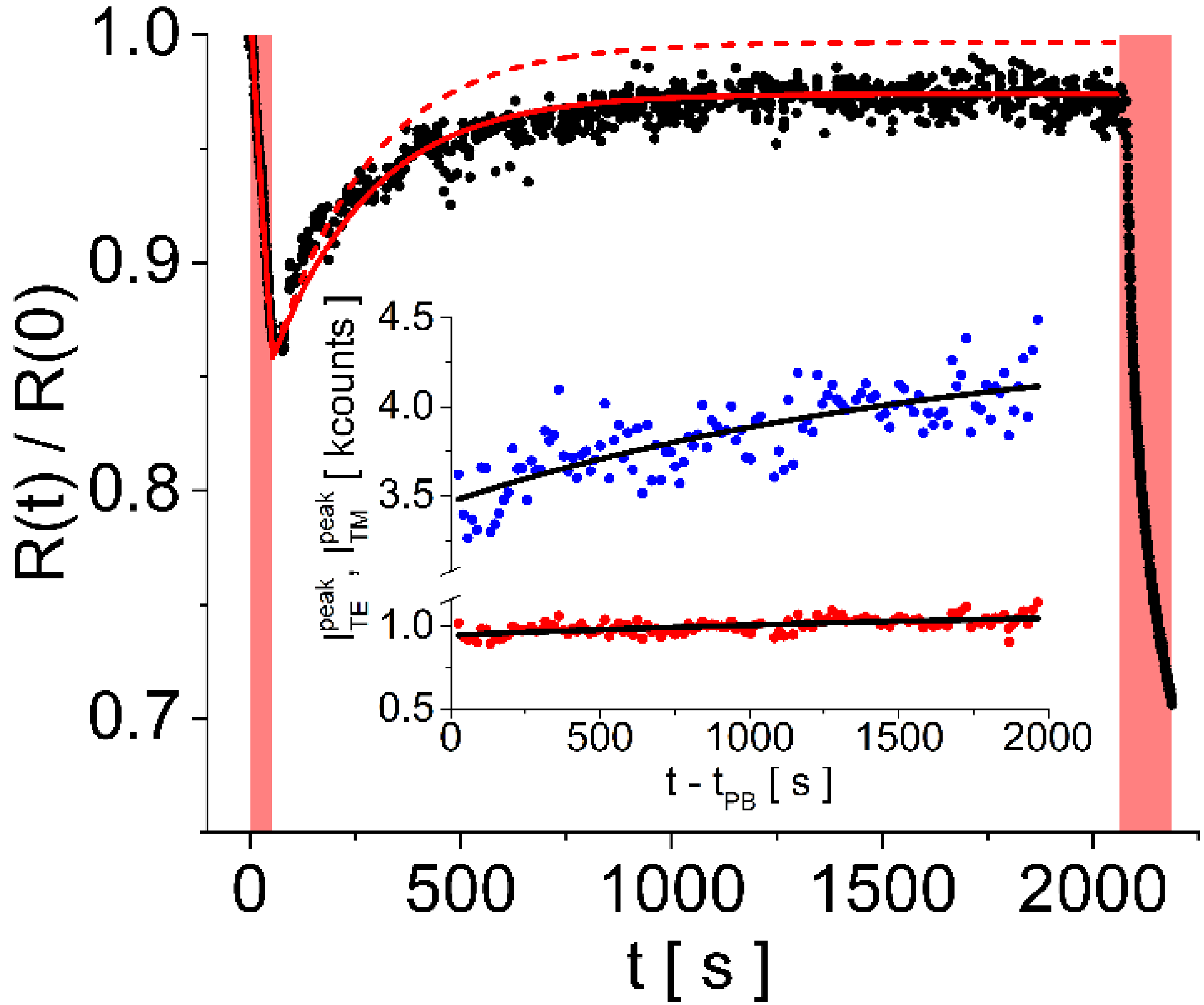

***Figure 4***. Temporal dependency of the normalized $R(t)/R(0)$ extracted from the experimental curves shown in Figure 3. The fits were obtained by the Eq. (1), when assuming that all residual active dipoles are either free to rotate (dashed) or partially immobilized (solid). The inset shows a magnified plot of the $I_{TE}^{peak}$ and $I_{TM}^{peak}$ signals shown in figure 3, limited to the fluorescence recovery phase. The last part of the measurement (second red region) was acquired under PB conditions.